# COVID-19 Infection Exposure to Customers Shopping during Black Friday

Braxton Rolle[1], and Ravi Kiran[2]

## Abstract

The outbreak of COVID-19 within the last two years has resulted in much further investigation into the safety of large events that involve a gathering of people. This study aims to investigate how COVID-19 can spread through a large crowd of people shopping in a store with no safety precautions taken. The event being investigated is Black Friday, where hundreds or thousands of customers flood stores to hopefully receive the best deals on popular items. A mock store was created, separated into several different shopping sections, and represented using a 2-D grid where each square on the grid represented a 5' × 5' area of the mock store. Customers were simulated to enter the store, shop for certain items, check out, and then leave the store. A percentage of customers were chosen to be 'infective' when they entered the store, which means that they could spread infection quantum to other customers. Four hours of time was simulated with around 6,000 customers being included. The maximum distance exposure could be spread (2'-10'), the minimum time of exposure needed to become infected (2 - 15 minutes), and the total percentage of customers who started as infective (1% - 5%) were all changed and their effects on the number of newly infected customers were measured. It was found that increasing the maximum exposure distance by 2' resulted in between a 20% to 250% increase in newly infected customers, depending on the distances being used. It was also found that increasing the percentage of customers who started as infective from 1% to 2% and then to 5% resulted in a 200% to 300% increase in newly infected customers.

**Keywords**: Air-borne illness; Social distancing; In-person shopping; Shopping malls; and Close contact exposure.

## 1.0 Introduction

The COVID-19 pandemic [1], has greatly impacted the lives of people worldwide and cost many lives. This has caused several nations and groups to come together and work towards the end goal of eliminating and mitigating the impacts of COVID-19 [2]. Some of the most common attempts at slowing or stopping the spread of COVID-19 were based on lowering or removing potential chances for the spread of the disease to occur between any two individuals since the disease spread most commonly from close contact with an infected person [3]. Some of these precautions included wearing masks while indoors, avoiding close contact or crowded areas of people (social distancing), and getting vaccinated as soon as possible [4], all of which were deemed effective for slowing the disease's spread [5][6][7]. Some restrictions did not work as well though. One study found that when the second set of lockdowns went in place in Toronto and Peel Canada, there was a significant increase in travel to surrounding regions for restaurants and grocery stores, meaning the lockdowns put in place were not as effective as they were meant to be [8].

---

[1] Undergraduate Researcher, Dept. of Computer Science, North Dakota State University, Fargo, ND 58105, email: braxton.rolle@ndsu.edu
[2] Associate Professor (Corresponding author), Dept. of Civil & Environmental Engineering, North Dakota State University, Fargo, ND 58105, email: ravi.kiran@ndsu.edu

The lifestyle changes did not come without consequences though. Due to the attempted avoidance of social contact by many people, small businesses struggled and either had to adapt or close down [9], and even some large industries that require close human contact suffered great losses and had to make major changes just to run the business [10]. The restaurant/hospitality industry has taken some of the hardest losses. One study found that over half of the people involved in their survey would not order out any longer, which is most likely due to the lasting impact of COVID-19 [11]. These changes did not just affect businesses and companies, however, like many places where people spent lots of time also had to make changes. School closures alone were estimated to inhibit the learning of over 100 million additional children [12]. Many workplaces were also affected with many either forced to go completely virtual [13] or in severe cases lay off many employees, resulting in record-high unemployment rates in the US [14]. Large events where many people would gather and be in close proximity of each other were almost entirely canceled. This included worldwide events such as the 2020 Olympic games [15] and Expo 2020 [16], which were both postponed a year. Even after these events took place again, many people were cautious about returning so quickly, especially when new cases were still appearing. One study found that in England, Germany, and Italy, football stadium attendance seemed to be affected by the previous days' COVID-19 confirmed case or death count [17].

Amidst the reopening of countries, many studies were created in analyzing, discussing, and identifying the potential changes that could be made in the physical architecture of cities and buildings to better prepare for another pandemic or outbreak. Some studies took the route of analyzing what elements of city infrastructure/architecture, called built environment, are potentially damaging to communities during COVID-19 [18][19]. One specific study proposes changes that could happen in large cities, which consider the current density and compact structure that many cities are aiming for and look to reduce this in the future [20]. Two studies took a targeted focus on public transport in the Netherlands and how to manage the COVID-19 capacity limits more effectively by modifying the routes the trains take to mitigate losses to revenue [21][22]. Some studies set their focus on building-specific solutions, rather than big city-wide changes. One of these studies uses a model to determine the correct locations in a building for air cleaners to be positioned to increase the air quality as much as possible [23]. Another of these studies used machine learning models to try to predict the air pollution concentrations in a commercial building over a set period, which would allow a smart air control system to correctly identify when and where the air quality was lacking [24]. These changes to city and building infrastructure are important because, in the case of another lockdown being required, people in the area still need access to essential buildings/services.

By mid to late 2020, the United States was in the process of reopening starting with things like schools, universities, and supermarkets [25]. This also meant that many large events started to take place in person again, such as Black Friday. Black Friday is a shopping holiday and takes place the day after Thanksgiving. It is one of the biggest shopping days in the United States, where over 30% of US population plan to partake. This number has decreased over the past couple of years [26], but the Black Friday weekend still amassed over 150 million shoppers in 2020, with over 58 million people still shopping in person [27] despite the recent COVID-19 pandemic. Surveys also showed that of the people that shopped online, over 57% had concerns over health and safety due to the pandemic [28], which shows how concerned people were despite the reopening. Within a month of the Black Friday 2020 shopping event, the COVID-19 delta variant was identified and quickly swept through England and the United States [29]. This variant was much more contagious, and the current vaccines had reduced efficiency against it [30]; however, most countries did not go into another lockdown similar to how they did when the first COVID-19 outbreak occurred.

With many shoppers eager to return to in-person shopping, and a far more contagious variants of COVID-19 still ravaging healthcare systems, Black Friday 2022 has the potential to cause a great spike in the COVID-19 cases in America. Although more and more people are sticking to online shopping every year, tens of millions of Americans are expected to attend in person. This study aims to find out the impact on new COVID-19 cases if customers choose to shop in person. The objective of this study is to simulate what will happen when infected and non-infected customers move through a mock store, and how the infection exposure will be affected by the distance between customers and how long the customers spend with each other.

## 2.0 Methodology

The number of total starting infective customers, the furthest distance that infection could be spread, the total amount of time someone could be exposed until they were deemed infected, and whether or not newly infected customers could spread infection were all parameters that were tested in this study.

### 2.1 Store Layout

The store layout in this study is not representative of any particular superstore, grocery store, etc., but follows a general structure that most of the North American stores do. The store is split into several large sections grouped by what would be sold there. For instance, the electronics section would contain all the store's video games, computers, movies, phones, etc. The areas selected in this study where more products are available for customers are areas that typically have more deals on Black Friday. For example, the Electronics section would have significantly more products available than the Produce section due to the higher interest in that area for Black Friday Deals [31]. The general store layout used for this study is shown in Figure 1. This store has around 120,000 ft$^2$ of floor space, which is similar to other superstores; Walmart averages around 178,000 ft$^2$ [32], Target averages around 130,000 ft$^2$ [33], and Kroger Combination stores average around 71,000 ft² [34]. This store is then discretized using a grid, which is used for the simulations conducted in this study.

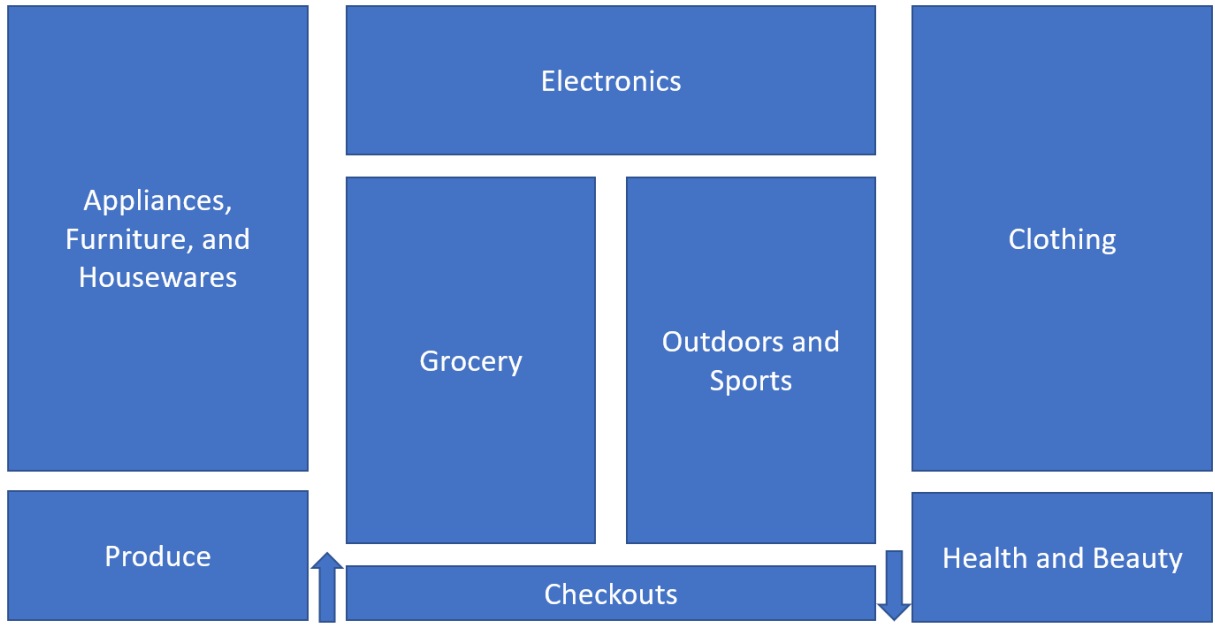

Figure 1: General mock store layout used in this study with sections labeled. (Arrows facing into the store entrance, and those facing away are exits)

The tests conducted were done using a two-dimensional grid. This study used a grid as a representation of the general superstore given in the previous section. This grid had dimensions of 60 units × 80 units, and each of these units represented a 5' × 5' area. The origin (0,0) position of the grid was in the bottom left corner of the store plan. Increasing the 'x' value moves to the right and increasing the 'y' value moves up. The shape of the discretizing units for the grid were squares. Each position on the grid could either be open or blocked. Agents (customers/ staff) can traverse through open positions but not through closed positions.

There were 34 positions selected throughout the store, shown in red in Figure 2, that were made to be Black Friday product locations that the customers would have to search for. There were also 42 positions, split into 21 groups of two and shown in blue in Figure 2, that were chosen as checkouts where the customers would go to purchase the products that they found. There are three exit and three entrance positions, colored green and yellow respectively, which is where all customers either leave or enter the store. All other positions are either open positions, where customers can walk, or closed positions, where they cannot. Closed positions are used to represent aisles in the store. This mock store is shown in Figure 2. Customers would traverse through this store by first entering, then moving to each product that they need and getting it, then going to the checkouts, and finally exiting.

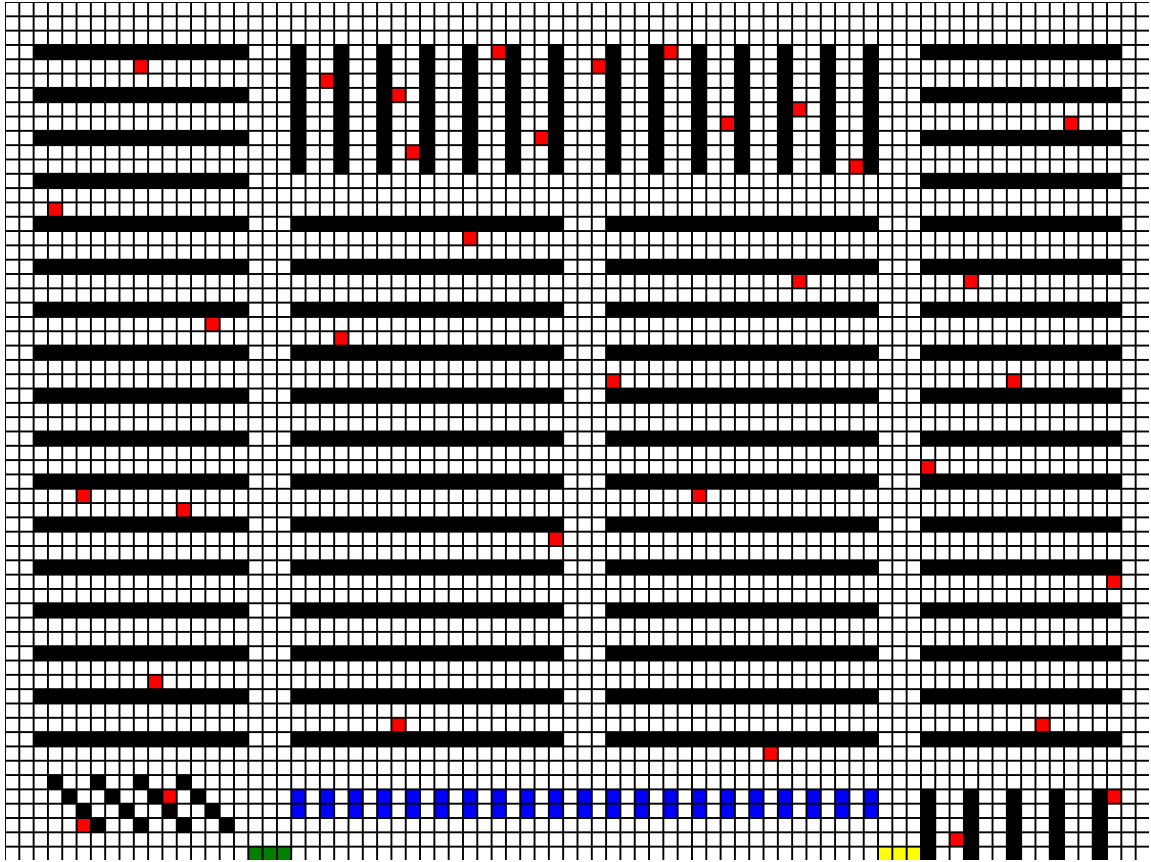

Figure 2: Discretization of the mock store layout using 5'X5' square grid (Red = Black Friday product, Blue = checkout, Black = aisle/blocked square, Green = entrance, Yellow = exit)

## 2.3 Navigation through the Store: Path-finding Algorithm

Customers are the primary agents that were used to traverse the store in this study. As stated before, customers enter the store, find their product, check out, and then leave the store. Before each customer enters the store, a number between three and eight is randomly chosen, which is the number of products that the customer will search for. Then, this same number of products are randomly selected, which they must traverse through the store to find. Once they are finished, they move towards the closest checkout and then to one of the exits. To simulate movement from thousands of customers to several points each, a pathfinding algorithm was used.

The pathfinding algorithm used in this study was the A* pathfinding algorithm. This was chosen due to its efficiency over other algorithms of the same type and ability to deliver the shortest path between two grid points [35]. It works by calculating three parameters ($F$, $G$, and $H$). The $G$ value is the calculated Euclidean distance between the starting (initial) position and the current position. This calculation is given in Eq 1. $(x_i, y_i)$ are the coordinates of the center of the initial position and $(x_c, y_c)$ are the coordinates of the center of the current position.

$$G = \sqrt{(x_i - x_c)^2 + (y_i - y_c)^2} \qquad \text{Eq 1}$$

The $H$ value is known as the heuristic and is the estimated distance from the current point to the target point. The estimation used was in Manhattan distance rather than Euclidean distance. This calculation is given in Eq 2. $(x_t, y_t)$ are the coordinates of the target position, $(x_c, y_c)$ are the coordinates of the current position and $|*|$ returns the absolute value of the input. The value is then multiplied by $l^* = 5'$ which is the length of each fundamental grid unit in feet.

$$H = (|y_t - y_c| + |x_t - x_c|) \times l^* \qquad \text{Eq 2}$$

The $F$ value is the calculation of the total overall distance needed to travel from each position to the target position. This calculation is given in Eq 3.

$$F = G + H \qquad \text{Eq 3}$$

The $F$ value is the primary parameter used in pathfinding, and the $G$ and $H$ values are only used to calculate it unless two $F$ values are the same, in which the $G$ value will be used to consider which position to choose. The A* pathfinding algorithm works by first calculating all of the $G$, $H$, and $F$ values around the starting position, and then selecting the position with the smallest $F$ value. The algorithm then checks all of the new potential positions and once again selects the position with the smallest $F$ value from any of the previously checked positions. This process is repeated until the target position is found. An example of a customer navigating through the store is shown below in Figure 3. Two examples of all customer positions at different time periods during a test is shown in Figure 4 (a) and (b).

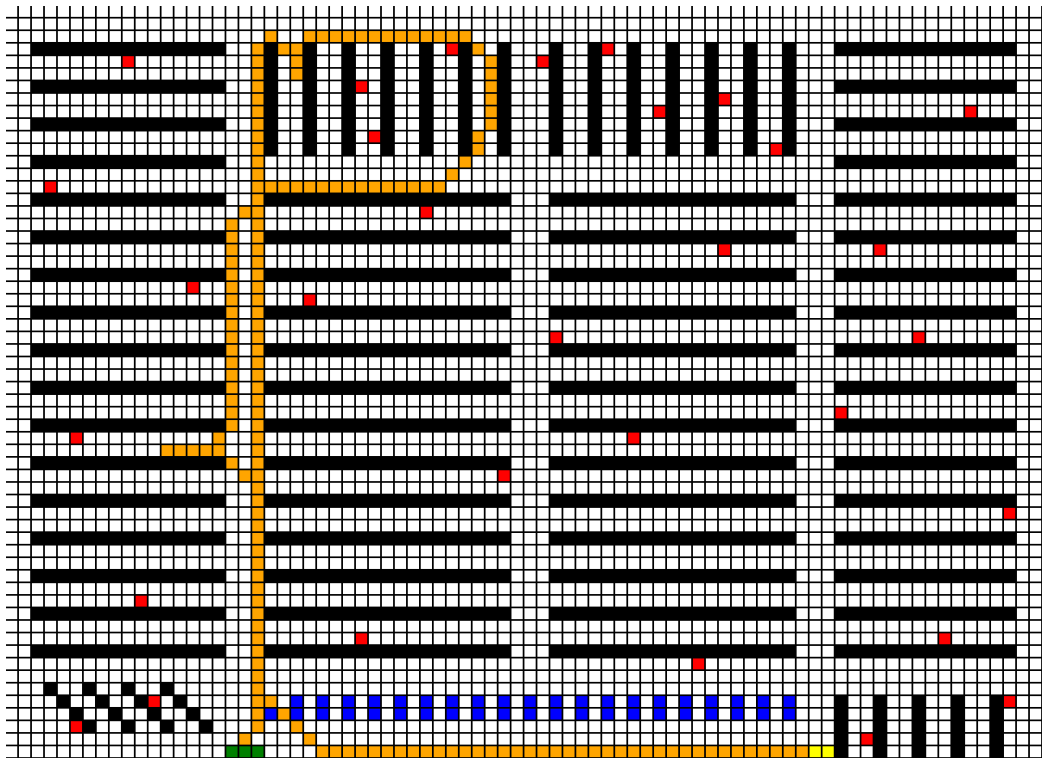

Figure 3: Example of a customer navigating through a store while shopping for three products
(Orange = customer path)

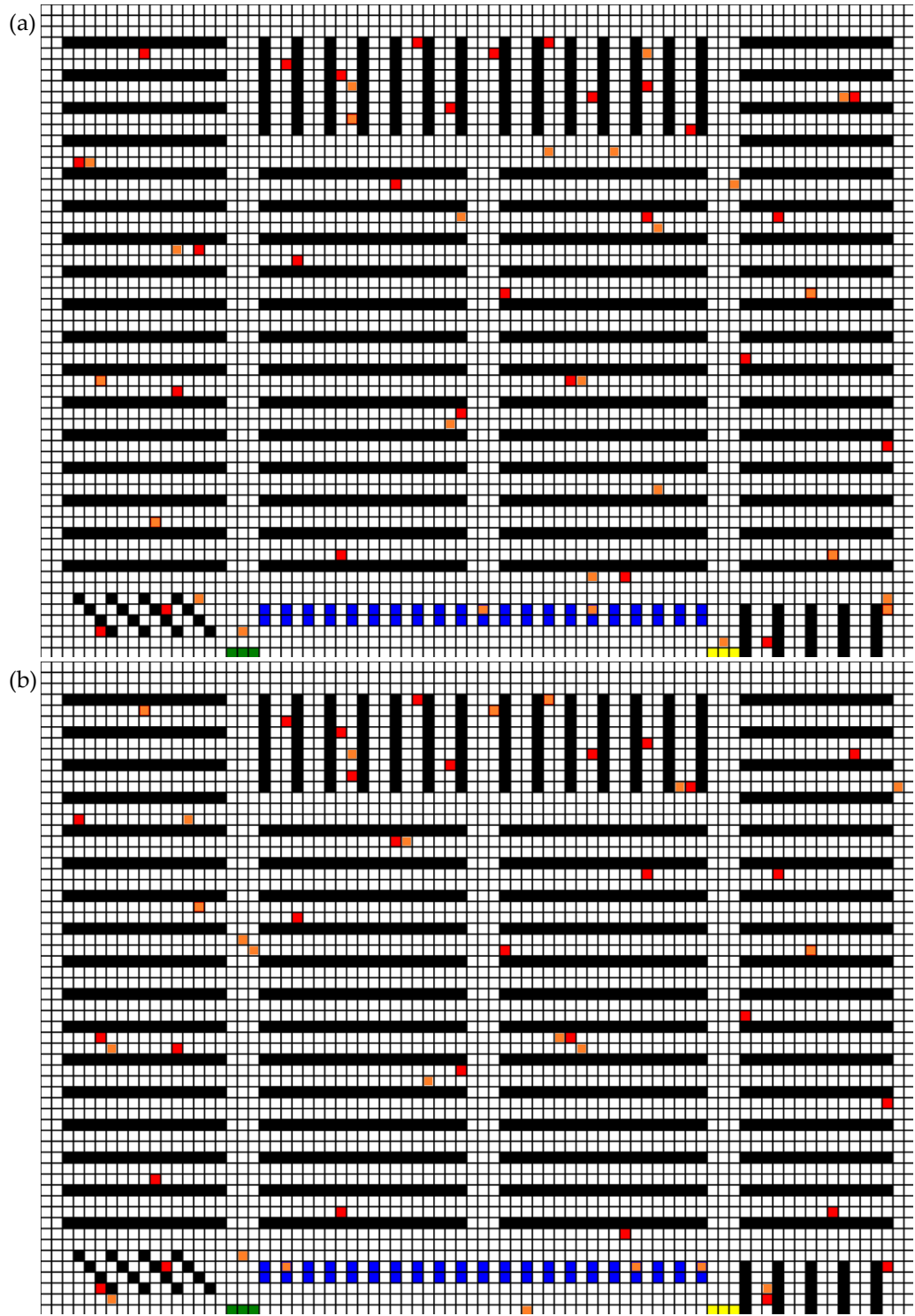

Figure 4: Typical locations of customers in the store at (a) 1 hour, and (b) 2 hours simulation period (Orange = Current position of a customer).

## 2.4 Simulation of Customers in the Shopping Area

All the simulated customers entered the store through the entrance, discretized by three squares. The customers will start at the closest product to the entrance and then move to the next closest product from their current position. Once they are in the same position as the product they were looking for, they will remove the product from their list and start moving towards the next product. Once they find all the products on their list, they will move towards the closest open checkout, and then to the exit, also discretized by three squares.

This was simulated for 6,000 customers over a four-hour time period. To simulate the sudden rush of customers on Black Friday, the store would start empty. The amount of customers entering the store for each period of time is shown in Table-1. The total amount of customers for each test using the estimation should be 6,000 but is less when recorded because only customers that leave the store during the four-hour period are counted, not those still in the store. Each customer was given an approximate walking speed of 5ft/s, which is close to the preferred walking speed of most people, 1.42m/s [36]. The walking speed used in this study is slightly higher because the Black Friday "rush" was taken into account, meaning customers would likely be moving faster than normal. The next section of this paper will introduce the infection exposure aspect of this study, which will use the simulation rules and details discussed.

Table 1: Number of customers entering the store for each period of time

| Time Period (minutes) | New Customers Introduced | Estimated Total |
|---|---|---|
| 0 – 30 | Two new customers every 1.5 seconds | 2400 new customers |
| 30 – 60 | One new customer every 1.5 seconds | 1200 new customers |
| 60 – 120 | One new customer every 3 seconds | 1200 new customers |
| 120 – 240 | One new customer every 6 seconds | 1200 new customers |

## 3.0 Tracking Infection Exposure During the Simulation

The primary purpose of this study is to measure infection exposure rates and the effect that the Black Friday rush of customers would have on the number of customers exposed to COVID-19. As customers entered the store simulation, a random percentage of them were selected as "infective", meaning that they would be able to spread the infection to other customers that are near them. This percentage of customers was between 1% - 5%, resulting in around 50 - 70 of the total customers being deemed "infective" from the start of the simulation, depending on the test being run.

Infection would be spread between an infective customer and a non-infective customer when they are within a set distance from each other. Distance is measured from the center of one position to the center of another. This distance would be 6' - 12' depending on the test being run. Each second that a non-infective customer spends near an infective customer adds a second to their total infection time. Once the total infection time for a customer reaches a threshold time, this customer would then be deemed infective. These newly infective customers would be the primary variable being tracked throughout the tests run in this study. An example of how infection exposure would be tracked is shown in Figure 5.

The grid shown in Figure 5 is a mock layout of how four different customers could be arranged and the distances between them are labeled. The blue positions are where the three non-infective customers would be standing, labeled Customer 1-3, and the red position is where the infective customer would be, labeled "Customer In". In this example, if the maximum exposure distance were 6', then Customer 1 would accrue infection exposure for as long as they stay the same or less distance from Customer In, and the other two non-infective customers would not accrue any infection exposure. If the maximum exposure distance were 8', then both Customer 1 and Customer 2 would accrue one second of infection exposure, but Customer 3 still would not.

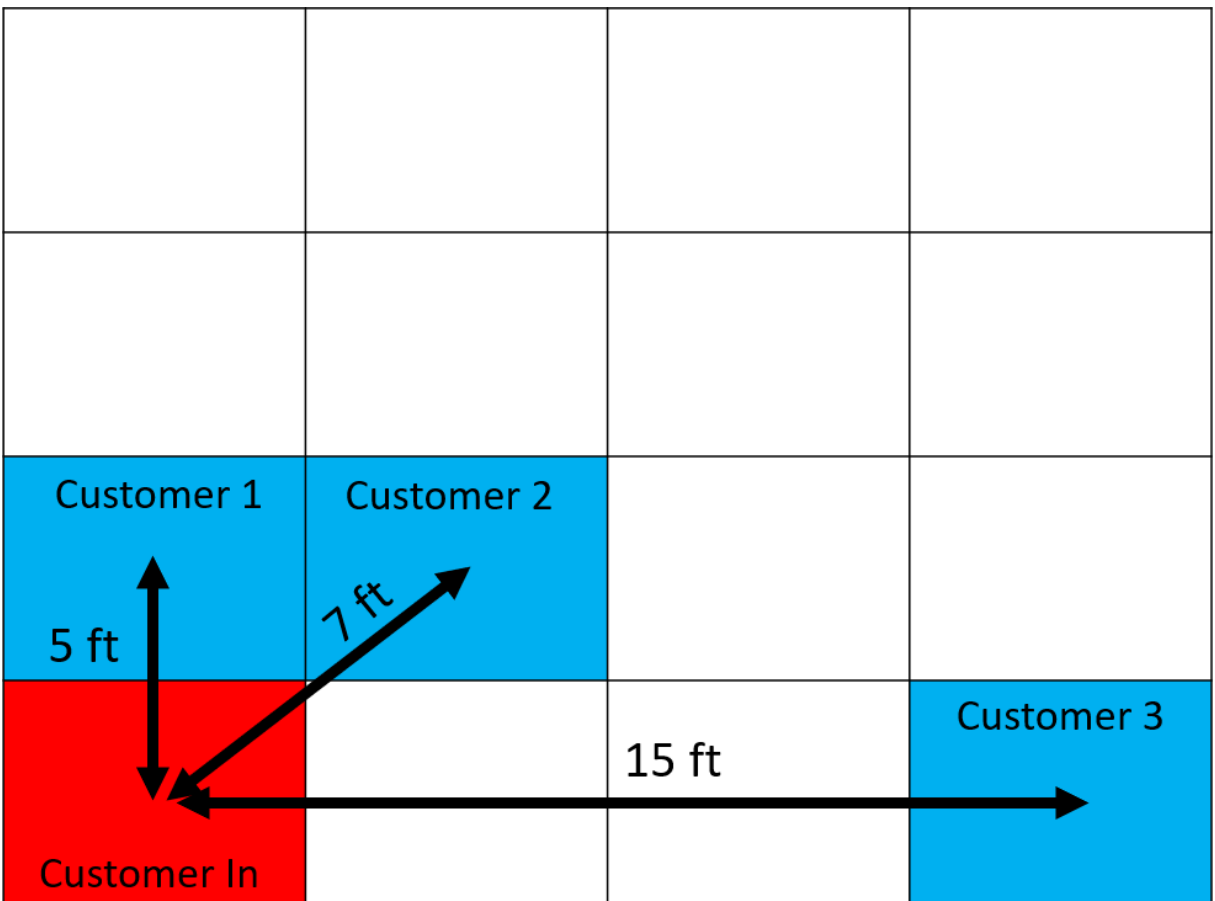

Figure 5: Example of how infection exposure is tracked, and how the exposure distance is measured (Grid unit size- 5'x5')

Two assumptions had to be made to run the simulations in this study. The first assumption is that customers could be partially exposed. This means that if a non-infective customer was exposed to one infective customer, walked away, and then was exposed to another infective customer, these separate instances of exposure would compound onto each other; A non-infective customer that was exposed to two separate infective customers for five total minutes would be treated the same as if they were exposed to one infective customer for five total minutes. The second assumption that had to be made was that newly infected customers could not infect non-infective customers unless this is considered as a parameter. Only customers that were selected as infective from the start of the simulation could contribute to the infection exposure of other non-infective customers.

## 4.0 Results of Exposure Tests

The four variables that were modified in this study were 1) the maximum exposure distance, 2) the exposure threshold as to when customers would be deemed infected, 3) the percentage of total customers selected as infective, and 4) whether or not newly infected customers could spread the infection to other non-infected customers. To start the test, 1% of total people were chosen as infective customers, and the maximum exposure distance was 6', which is in accordance with the U.S. Center for Disease Control's (CDC) guidelines of COVID-19 close contact [37]. The time threshold used was either 2 minutes, 5 minutes, 10 minutes, or 15 minutes. The CDC guidelines say that 15 minutes of exposure over the course of 24 hours is considered close contact, but for these tests, many customers would only be in the store for 15-30 minutes total, so it would be nearly impossible for them to accrue enough infection to become infected, which is why the lower time thresholds were also tested considering the worst-case scenario where a new highly infective variant could emerge. Table 2 shows the results from the first set of tests. For each test, the number of starting infective customers will vary because each customer is given a 1% chance to be deemed infective when they enter the store. This leads to the number of starting infective being around the 1% range, but not exactly 1% of total customers. Similarly, the number of total customers will always be close to 6000, but only customers that enter and exit the store over the 4-hour period will be counted in the total. This means that customers that enter the store but do not exit during the time period will not be counted in the total, leading to the slight variance shown in the tables.

Table 2: Results of the first set of tests (6ft max exposure distance, 1% originally infected customers, no newly infected exposure)

| Max Exposure Distance | 6' | | | |
|---|---|---|---|---|
| Time Threshold (minutes) | 15 | 10 | 5 | 2 |
| Starting Infective Customers | 58 | 52 | 65 | 47 |
| **Newly Infected Customers** | **0** | **0** | **0** | **30** |
| Total Customers | 5941 | 5926 | 5923 | 5922 |

As can be seen in Table-2, three of the tests resulted in no new infected customers, and the fourth test resulted in around .5% of total customers becoming infective. The biggest factors that influence these small numbers of newly infected customers are the large size of the store combined with the small amount of total infected customers, and the short exposure distance. The mock store used has a lot of open space within it, and when only 47-65 customers are inside it over the course of four hours, there is a very small chance that any other customer will be within the maximum exposure distance of them for even a couple minutes. At most, there would be 2-3 infective customers in the store at once, which results in very little infection spread.

The exposure distance being 6' also results in a minuscule amount of infection spread occurring. Using the grid is shown earlier, which has 5' × 5' squares as its fundamental unit, in order for a person to accrue infection exposure, they would have to be within the same 5' × 5' square as another customer, or in the square directly East, West, North, or South of them. This is shown in Figure 6, with the red square being the infective customer and the orange squares being the positions that are vulnerable to infection exposure; The position with the infective customer also is vulnerable to infection exposure. Being in one of the four squares in any of the diagonal directions would not result in exposure being spread because they would be roughly 7' apart. This means that for each infective customer, there are only five possible positions that another customer could be in for them to accrue infection exposure, which when divided by the total number of possible positions, results in only about a .1% total chance that a randomly selected position would be exposed to infection.

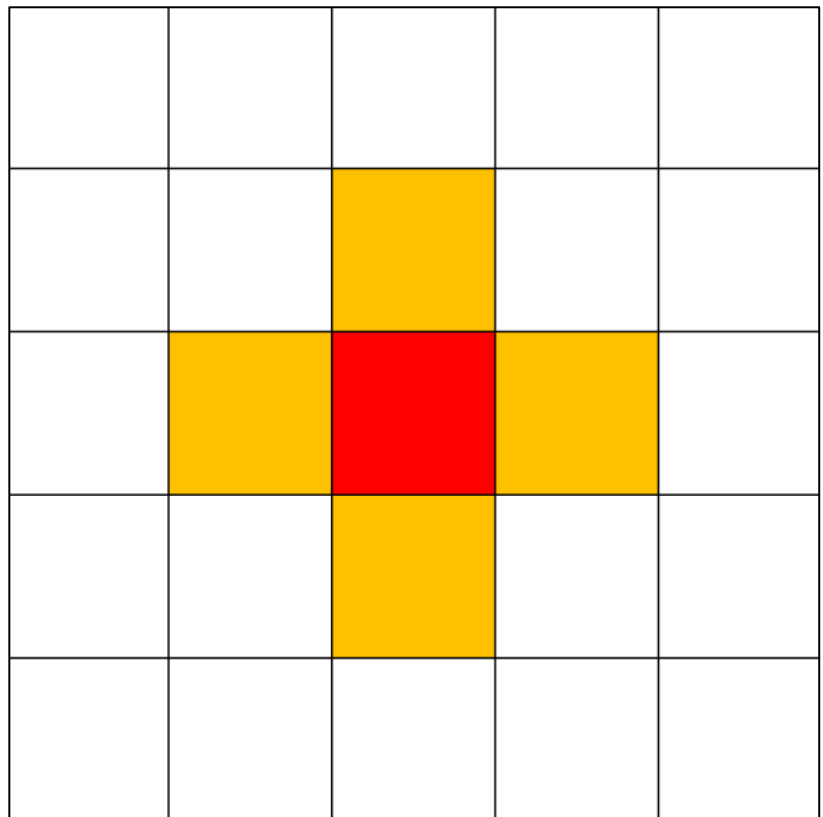

Figure 6: Heatmap of a maximum exposure distance of 6' (Red = infective customer, Orange = Vulnerable to Exposure, 5'x5' unit size).

As stated earlier, the simulation of a "Black Friday Rush" also limits the amount of time that customers are in the store. Most customers are not going to be browsing or shopping how they normally would; Instead, they would be moving quickly throughout the store and exciting as quickly as possible. This makes it difficult for 15 minutes of exposure to occur to any customer, as the odds that they are near another customer, much less an infected customer, for an extended time are very slim.

## 4.1 Measuring the Effects of Increasing Maximum Exposure Distance

The first expansion of the tests was to test how increasing the maximum exposure distance would affect the total number of newly infected customers. The original maximum exposure distance used was 6', which is in accordance with the CDC guidelines for close contact; However, due to the lack of newly infected customers in the first test, larger distances were used to see how it would affect total exposure. The next two distances used were 8' and 10'. These distances were chosen because of the natural increase of total possible positions around each infective customer. This concept is displayed in Figure 7 using heat maps of infection exposure with a maximum distance of 8' (a) and 10' (b).

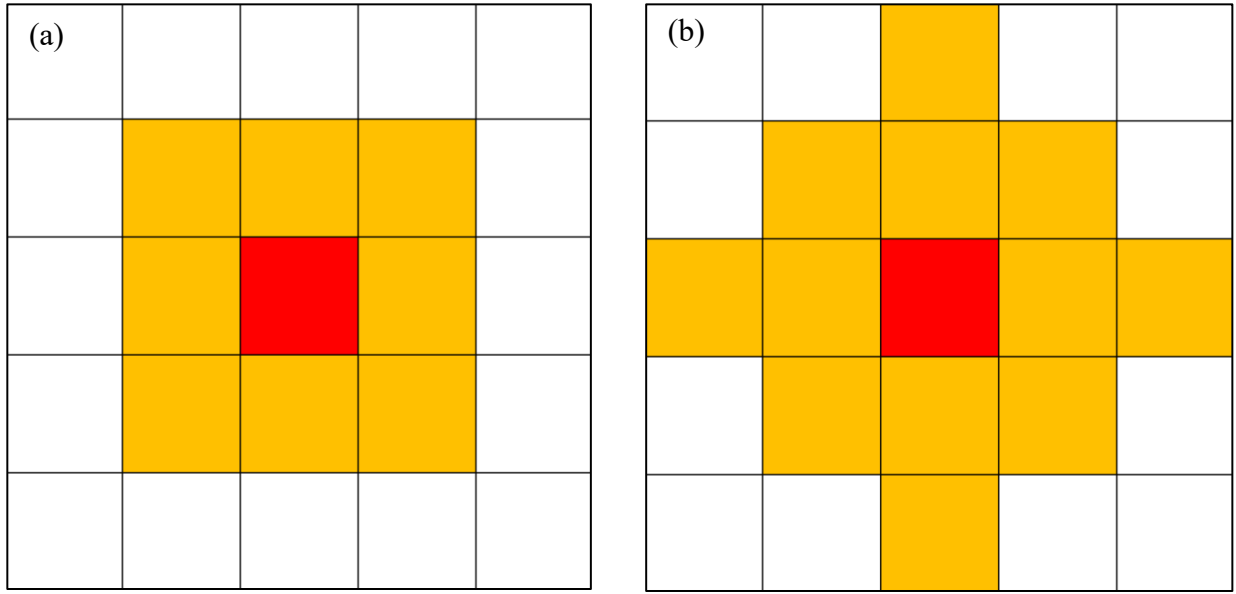

Figure 7: Heatmap of infection exposure using maximum distance of (a) 8', (b) 10' (Red = Infective Customer, Orange = Vulnerable to Exposure, 5'x5' unit size)

These heatmaps display the positions around an infective customer that would result in infection exposure for another customer. Each maximum distance interval used has four more vulnerable positions than the one prior: A maximum distance of 6' results in five vulnerable positions, a maximum distance of 8' results in nine vulnerable positions, and a maximum distance of 10' results in thirteen vulnerable positions. The results of the tests using a maximum distance of 8' are shown in Table 3.

Table 3: Results of tests using a maximum distance of 8' (1% total infected, no newly infected exposure)

| Max Exposure Distance | 8' | | | |
|---|---|---|---|---|
| Time Threshold (minutes) | 15 | 10 | 5 | 2 |
| Starting Infective Customers | 49 | 65 | 64 | 55 |
| **Newly Infected Customers** | **0** | **0** | **3** | **111** |
| Total Customers | 5920 | 5930 | 5923 | 5931 |

As seen in Table 3, the increased maximum exposure distance affects the number of customers infected. To start, using a time threshold of five minutes now results in three customers becoming infective during the simulation compared to the first test which resulted in no customers becoming infective. Using a time threshold of two minutes also resulted in 111 newly infected customers, a little less than 2% of the total customers in the store compared to the .5% seen in the prior test. The tests using time thresholds of 10 minutes and 15 minutes still resulted in no customers becoming infected. The results of the tests using a maximum distance of 10' are shown in Table 4.

Table 4- Results of tests using a maximum distance of 10' (1% total infected, no newly infected exposure)

| Max Exposure Distance | 10' | | | |
|---|---|---|---|---|
| Time Threshold (minutes) | 15 | 10 | 5 | 2 |
| Starting Infective Customers | 48 | 53 | 63 | 61 |
| **Newly Infected Customers** | **0** | **0** | **5** | **135** |
| Total Customers | 5929 | 5930 | 5926 | 5924 |

The difference between a maximum exposure distance of 10' and 8' was observed to be much less than the difference between 8' and 6'. Still, no customers were infected during the 15- and 10-minute thresholds. The 5-minute threshold resulted in five customers becoming infected, which is an increase of 2 over the prior test, and the 2-minute threshold resulted in 135 customers becoming infected, which is a little over 2% of the total customers and 24 more than the prior test.

The final maximum exposure distance value tested was 12'. Unlike the previous three distance values, this distance has 8 more vulnerable positions compared to the previous distance interval, compared to the others which only have four. This results in 21 total positions that are vulnerable to infection exposure for each infective customer: Over 4 times the amount from the first set of tests. The heatmap for a 12' maximum exposure distance is shown in Figure 8. The results from this set of tests are shown in Table 5.

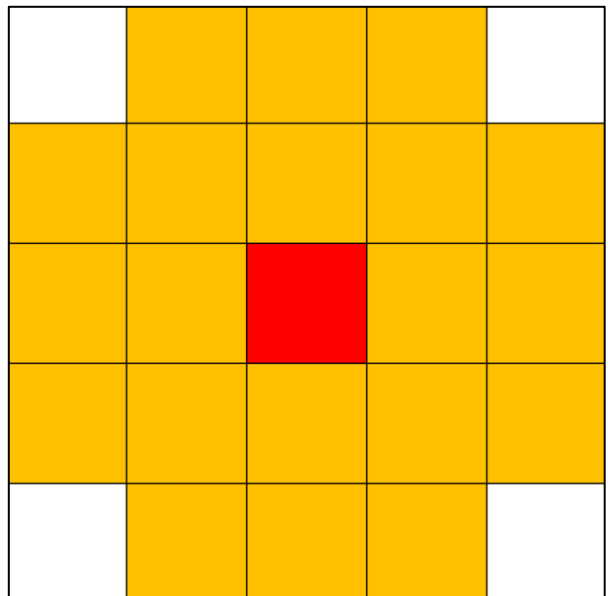

Figure 8: Heatmap of infection exposure using a maximum distance of 12' (Red = Infective customer, Orange = vulnerable to Exposure, 5'x5' unit size)

Table 5: Results of tests using a maximum distance of 12' (1% total infected, no newly infected exposure)

| Max Exposure Distance | 12' | | | |
|---|---|---|---|---|
| Time Threshold (minutes) | 15 | 10 | 5 | 2 |
| Starting Infective Customers | 54 | 62 | 59 | 59 |
| **Newly Infected Customers** | **0** | **0** | **10** | **222** |
| Total Customers | 5899 | 5937 | 5913 | 5927 |

These tests still have no newly infected customers at the 15- and 10-minute thresholds but have double the newly infected at the 5-minute threshold when compared to the prior test and have about 64% more newly infected customers at the 2-minute threshold. These large jumps are to be expected when considering how much more area around each infective customer is vulnerable to infection exposure when compared to the prior tests. From these tests, it can be observed that the higher the maximum distance for infection spread is, the more customers will not only be exposed but infected by the starting infective customers. It can also be seen that decreasing the amount of time needed to fully "infect" new customers will result in higher levels of infection. The total number of customers and number of starting infective customers vary slightly from test to test, but do not seem to affect the results of what has been observed so far.

**4.2 Measuring the Effects of Increasing Originally Infective Customers**

After observing the effects that an increased maximum exposure distance would have on the number of newly infected customers, the next element of the original test that was further investigated is the number of starting infective customers. Up to this point, all tests have used around 1% of total customers as starting infective. The percentages that were tested next are 2% and 5%. First, 2% of total customers as starting infective was tested. This resulted in around 100-140 customers entering the store as infective and able to spread infection exposure. The same test setup was used as the previous tests. All four max distances were used (12', 10', 8', 6'), and all the four time thresholds (15, 10, 5, 2 minutes) were used resulting in 16 total tests run using the 2% of the total as starting infective. The results from this test are split between Table 6, which shows the 12' and 10' maximum distance results, and Table 7, which shows the 8' and 6' maximum distance results. All these tests still do not allow newly infected customers to spread the infection to other customers.

Table 6: Results of tests using a maximum distance of 12' and 10' (2% total infected, no newly infected exposure)

| Max Distance | 12' | | | | 10' | | | |
|---|---|---|---|---|---|---|---|---|
| Threshold (mins) | 15 | 10 | 5 | 2 | 15 | 10 | 5 | 2 |
| Starting Infective | 127 | 119 | 138 | 109 | 124 | 104 | 124 | 130 |
| **Newly Infected** | **0** | **2** | **40** | **516** | **0** | **0** | **21** | **301** |
| Total Customers | 5929 | 5933 | 5932 | 5914 | 5899 | 5922 | 5941 | 5934 |

Table 7: Results of tests using a maximum distance of 8' and 6' (2% total infected, no newly infected exposure)

| Max Distance | 8' | | | | 6' | | | |
|---|---|---|---|---|---|---|---|---|
| Threshold (mins) | 15 | 10 | 5 | 2 | 15 | 10 | 5 | 2 |
| Starting Infective | 111 | 131 | 125 | 118 | 109 | 128 | 118 | 124 |
| **Newly Infected** | **0** | **0** | **7** | **249** | **0** | **0** | **1** | **108** |
| Total Customers | 5931 | 5903 | 5941 | 5904 | 5937 | 5941 | 5933 | 5933 |

At this point, changes can already be seen in the number of newly infected customers when compared to the prior tests. The test ran using a maximum distance of 12' and a 10-minute time threshold resulted in two customers becoming infected, which is the first test using a 10-minute threshold that resulted in newly infected customers. The 6' and 5-minute threshold also resulted in one customer becoming infected, which is the first test ran at those parameters (6', 5 minutes) that resulted in a newly infected customer. Taking a more general approach when comparing the results, all the newly infected customer values are 100%-300% higher than their counterpart tests ran previously; i.e., the 12' and 15-minute threshold test shown in Table 6 would correlate with the 12' and 15-minute threshold test shown in Table 5 and so on.

After the prior tests concluded, the number of total customers chosen as starting infectives was increased to 5%, and the same set of tests was run again with the only change being the number of starting infectives. The number of starting infective customers ranged from about 259-292. The results from this test are split between Table 8, which shows the 12' and 10' maximum distance results, and Table 9, which shows the 8' and 6' maximum distance results. All these tests still do not allow newly infected customers to spread the infection to other customers.

Table 8: Results of tests using a maximum distance of 12' and 10' (5% total infected, no newly infected exposure)

| Max Distance | 12' | | | | 10' | | | |
|---|---|---|---|---|---|---|---|---|
| Threshold (mins) | 15 | 10 | 5 | 2 | 15 | 10 | 5 | 2 |
| Starting Infective | 288 | 292 | 278 | 267 | 277 | 268 | 274 | 283 |
| **Newly Infected** | **0** | **4** | **137** | **1853** | **0** | **2** | **66** | **1014** |
| Total Customers | 5933 | 5926 | 5918 | 5922 | 5938 | 5935 | 5941 | 5919 |

Table 9: Results of tests using a maximum distance of 8' and 6' (5% total infected, no newly infected exposure)

| Max Distance | 8' | | | | 6' | | | |
|---|---|---|---|---|---|---|---|---|
| Threshold (mins) | 15 | 10 | 5 | 2 | 15 | 10 | 5 | 2 |
| Starting Infective | 274 | 267 | 269 | 259 | 280 | 277 | 291 | 288 |
| **Newly Infected** | **0** | **0** | **34** | **793** | **0** | **0** | **5** | **326** |
| Total Customers | 5911 | 5907 | 5928 | 5930 | 5941 | 5898 | 5927 | 5931 |

Similar results can be seen from these tests as the prior ones. The number of newly infected customers increased by about 100%-300% again in all tests where a customer was infected. The number of newly infected customers reached all the way up to 31% on one of the tests, meaning that out of all the customers to enter the store, over 31% of them left the store infected. All these tests display that increasing the number of customers that enter the store infected will subsequently raise the total number of newly infected customers at distances and time thresholds.

### 4.3 Influence of Newly Infected becoming Infective

At this point, several variables used throughout the tests have been expanded on and observed; However, the original assumptions that had to be made have not been investigated at all. The two assumptions are that infective customers could spread partial exposure to other customers and that newly infective customers could not spread the infection to other customers. The first assumption is necessary for receiving any useful data out of this simulation, as it is next to impossible that any customer would be within exposure distance of an infected customer for even two minutes, which would result in no newly infected customers for all tests. The second assumption, however, can be reversed and investigated.

The next set of tests ran to use the same four maximum exposure distances (12', 10', 8', and 6'), the same four-time thresholds (15, 10, 5, and 2 minutes), and 1% of total customers as starting infective. The difference between these tests and the prior ones is that newly infected customers can spread the infection to other non-infected customers. This change should not affect any tests where no new customers were infected, but it should increase the number of newly infected customers in all other tests. The results from this test are split between Table 10, which shows the 12' and 10' maximum distance results, and Table 11, which shows the 8' and 6' maximum distance results.

Table 10: Results of tests using a maximum distance of 12' and 10' (1% total infected, newly infected exposure allowed)

| Max Distance | 12' | | | | 10' | | | |
|---|---|---|---|---|---|---|---|---|
| *Threshold (mins)* | 15 | 10 | 5 | 2 | 15 | 10 | 5 | 2 |
| *Starting Infective* | 52 | 52 | 57 | 59 | 57 | 51 | 49 | 56 |
| **Newly Infected** | **0** | **0** | **105** | **2609** | **0** | **0** | **34** | **987** |
| *Total Customers* | 5899 | 5932 | 5931 | 5923 | 5935 | 5897 | 5891 | 5927 |

Table 11: Results of tests using a maximum distance of 8' and 6' (1% total infected, newly infected exposure allowed)

| Max Distance | 8' | | | | 6' | | | |
|---|---|---|---|---|---|---|---|---|
| *Threshold (mins)* | 15 | 10 | 5 | 2 | 15 | 10 | 5 | 2 |
| *Starting Infective* | 52 | 54 | 60 | 63 | 59 | 57 | 54 | 56 |
| **Newly Infected** | **0** | **0** | **1** | **881** | **0** | **0** | **0** | **302** |
| *Total Customers* | 5944 | 5929 | 5937 | 5935 | 5954 | 5908 | 5886 | 5940 |

The results from these tests came out mostly as expected. The tests that previously had no newly infected customers result from them still did not have any newly infected customers in this set of tests. The tests that did previously have newly infected customers, however, were increased significantly, except for one. The one test that did not have more newly infected customers than its corresponding test from before was the 8' and 5-minute threshold test. The most likely reason for this is that there was still such a small number of newly infected customers that the change in the assumption was not able to affect the amount of exposure spread, resulting in the newly infected customer number being almost the exact same as before. All the other tests had increased by up to 900% when compared to the prior tests.

These results show that assuming newly infected customers can infect other customers would drastically increase the number of customers infected. Although this assumption does not align with the current base of knowledge there is on COVID-19, meaning that the disease is not contagious as soon as someone is exposed, it allows us to see the consequences that would happen if another similar disease would appear.

## 5.0 Limitations and Scope for Further Studies

The simulations, tests, and other research done through this study has limitations when looking for real-world application. First, although the tests conducted in this study could be applied to most airborne illnesses, the primary emphasis is on the transmission of the COVID-19 delta/Omicron variants, which are relatively new. The information currently known about COVID-19 could change drastically over time, especially as new variations arise. A change in how COVID-19 is spread would completely alter the results found in this study, which means that this study would have to be redone/revised before the data could be used. Second, the infection spread measured in this study's tests is not fully representative of how we currently understand COVID-19 to spread. The approach used in this study to measure infection spread is similar to how it would be in a real-world scenario, only simplified. Expanding on how the infection spread is measured could alter the results that were found. Some expansions that would make the simulation more realistic include limiting the number of customers in the store, utilizing masks and social distancing regulations, establishing separate hours for vulnerable and elderly patients, and using temperature screening technology to detect potentially sick customers before they enter the doors.

Other factors that must be considered are the size of the grid units used, the size and layout of the mock store used, the number of total customers in each simulation, and how the customers move through the store and interact with other customers. The current grid units that were used in this study's tests were a representation of a 5' × 5' area of the mock store. Changing this size, either increasing or decreasing, could alter the paths that customers take, subsequently changing the distance between them and the amount of infection exposure they would accrue. The layout and size of the mock store are another set of variables that could alter the results found in this study. Moving the location of the products that customers could find or changing how the aisles are laid out would once again change the paths that the customers could take and alter how much infection exposure they would accrue. Increasing the size of the store while maintaining the same number of customers would most likely reduce the number of people that would become infected because there would be more overall space for customers; Reducing the size would most likely cause the inverse. Changing the number of customers would likely follow the pattern that changing the store size would, except inverted. Increasing the number of customers while maintaining store size would likely result in more infected customers and reducing the number of customers would likely result in less infected customers.

The way customers interact both with the store and with other customers plays the biggest role in how much they are exposed to infection. This study is limited by the fact that all customers are treated the same and act the same within the simulation, which is not representative of all people who shop on Black Friday. Everyone has different methods of shopping, walking speeds, and preferences on how to interact with other customers, all of which would affect how likely they are to deal with infection exposure. Similarly, some people may be susceptible to infection from COVID-19 than others, which is not considered in this study.

Despite all the limitations discussed prior, the results found in this study have many potential real-world applications and implications. First, for implications, the results found demonstrate how quickly COVID-19 can spread within large groups of people. A couple of infected people quickly can expose the dangerous virus to hundreds, if only partially at a time. Social distancing procedures previously used in shopping centers like one-way aisles, 6' spacing marks in checkout lines, and limited customers in a store at a time are all things that could be implemented much more frequently in order to reduce infection exposure. The results found could also encourage stores to provide more virtual opportunities for customers to shop and reap the benefits of Black Friday without having to risk their health in the process.

The results found and methodology used in this study could be applied to other areas of research and in the ongoing fight against COVID-19. This study could be used as a basis to expand upon, which could include things like more realistic customer interactions/movement and more realistic infection exposure. It could also be expanded to incorporate the current techniques used to limit exposure and spread; This would include vaccinations, masks, and store procedures, all of which have an impact on how much people will be exposed to COVID-19. Another application of this work would be to remove the Black Friday implications to see how COVID-19 would be transmitted in an average store day. Black Friday only occurs once every year, so seeing the results of an average day would be much more insightful in finding how to limit exposure in the long term. This application could also measure many more in-depth interactions between customers over a longer time period. The current study is limited in time to the typical Black Friday 'rush' but measuring a normal day at a shopping center would allow for a much longer simulation to occur.

This study could also be expanded beyond the scope of a shopping center and implemented into other venues and locations. For example, the spread of COVID-19 could be measured inside of a workplace using similar techniques to the ones that were used in this study. Concerts, schools, and sports games are all other places that could be investigated as to how much infection exposure would be expected to happen. This would allow for a much greater range of data to be collected, as well as an insight into where the most exposure occurs and how to stop it.

## 6.0 Conclusions

The important conclusions of this study are:

1. A simulated rush of customers during the Black Friday inside a large shopping center where 1% of customers are infected, will not result in any other customers becoming infected in accordance with the CDC's standards of close contact exposure. Customers will experience exposure to COVID-19 but will not be deemed infective.
2. Decreasing the total time of exposure needed for a customer to be deemed infective will result in more and more customers becoming infected. The time was decreased from 15 minutes to 10, 5, and 2. The further the time needed to infect is decreased, the more customers that will become infected over the course of the simulation, up to a 1000% increase in some tests.
3. Increasing the maximum distance that exposure can occur over will result in both more overall infection being spread and customers becoming infected. The distance intervals used were 12', 10', 8', and 6', with 6' being the standard. The number of new customers becoming infected increases by up to 250% when comparing the same tests between intervals of distance.
4. Increasing the starting percentage of customers who are infected will result in more new customers becoming infected. The original percentage of customers that were infective from the start of the simulation was 1%, and this was increased to 2% and 5%. At each interval, the number of new customers increased by around 200-300%.
5. Assuming that newly infected customers will be able to spread the infection to other customers drastically increases the number of total new customers that will become infected. The increase in total newly infected customers was over 1000% in some tests and only a little under in the other tests.
6. Limiting the number of infected persons entering the store through temperature screenings, practicing social distancing, and having a lower number of shoppers at any given time in the store will lower the infection exposure even during rush hours.

## Author Contributions

Conceptualization, Ravi Kiran; Formal analysis, Braxton Rolle; Funding acquisition, Ravi Kiran; Investigation, Braxton Rolle; Methodology, Ravi Kiran; Software, Braxton Rolle; Supervision, Ravi Kiran; Validation, Braxton Rolle; Visualization, Braxton Rolle; Writing – original draft, Braxton Rolle; Writing – review & editing, Ravi Kiran.

announcements/expo-dubai-2020-2/bie-general-assembly-officially-approves-expo-2020-dubai-date-change

17. Reade, J. J., & Singleton, C. (2020). Demand for public events in the COVID-19 pandemic: a case study of European football. Https://Doi.Org/10.1080/16184742.2020.1841261, 21(3), 391–405. https://doi.org/10.1080/16184742.2020.1841261

18. Frumkin H. 2021. COVID-19, the built environment, and health. Environ Health Perspect129(7):75001, PMID: 34288733, 10.1289/EHP8888.

19. Ma, S., Li, S. & Zhang, J. Diverse and nonlinear influences of built environment factors on COVID-19 spread across townships in China at its initial stage. Sci Rep 11, 12415 (2021). https://doi.org/10.1038/s41598-021-91849-1

20. Cheshmehzangi, A. (2021). Revisiting the built environment: 10 potential development changes and paradigm shifts due to COVID-19. *Journal of Urban Management*, *10*(2), 166–175. https://doi.org/10.1016/J.JUM.2021.01.002

21. de Weert Y, Gkiotsalitis K. A COVID-19 Public Transport Frequency Setting Model That Includes Short-Turning Options. Future Transportation. 2021; 1(1):3-20. https://doi.org/10.3390/futuretransp1010002

22. Gkiotsalitis, K. (2021). A model for modifying the public transport service patterns to account for the imposed COVID-19 capacity. Transportation Research Interdisciplinary Perspectives, 9, 100336. https://doi.org/10.1016/J.TRIP.2021.100336

23. Dai, H., & Zhao, B. (2022). Reducing airborne infection risk of COVID-19 by locating air cleaners at proper positions indoor: Analysis with a simple model. *Building and Environment*, *213*, 108864. https://doi.org/10.1016/J.BUILDENV.2022.108864

24. Mohammadshirazi, A., Kalkhorani, V. A., Humes, J., Speno, B., Rike, J., Ramnath, R., & Clark, J. D. (2022). Predicting airborne pollutant concentrations and events in a commercial building using low-cost pollutant sensors and machine learning: A case study. Building and Environment, 213, 108833. https://doi.org/10.1016/J.BUILDENV.2022.108833

25. *This is where all 50 states stand on reopening - CNN.com*. (2020). Retrieved November 12, 2021, from https://www.cnn.com/interactive/2020/us/states-reopen-coronavirus-trnd/

26. • *Black Friday: consumers planning to shop U.S. 2020 | Statista*. (2020). Retrieved November 12, 2021, from https://www.statista.com/statistics/247010/americans-shopping-on-black-friday/

27. *Black Friday History and Statistics | BlackFriday.com*. (2020). Retrieved November 12, 2021, from https://blackfriday.com/news/black-friday-history

28. *78 Black Friday Statistics You Must Read: 2020/2021 Market Share & Data Analysis - Financesonline.com*. (2020). Retrieved November 12, 2021, from https://financesonline.com/black-friday-statistics/

29. Katella, K. (2021). *5 Things To Know About the Delta Variant > News > Yale Medicine*. https://www.yalemedicine.org/news/5-things-to-know-delta-variant-covid

30. Shiehzadegan, S., Alaghemand, N., Fox, M., & Venketaraman, V. (2021). Analysis of the Delta Variant B.1.617.2 COVID-19. *Clinics and Practice 2021, Vol. 11, Pages 778-784*, *11*(4), 778–784. https://doi.org/10.3390/CLINPRACT11040093

31. *Adobe Holiday Shopping Forecast*. (2021). Adobe. https://www.adobe.com/marketing/pdf-page.html?pdfTarget=aHR0cHM6Ly9idXNpbmVzcy5hZG9iZS5jb20vY29udGVudC9k
21